# Influence of nano-hole defects and their geometric arrangements on the superfluid density in atomically thin single crystals of indium superconductor

Mengke Liu[1], Hyoungdo Nam[1], Jungdae Kim[1,2], Gregory A. Fiete[3,4], Chih-Kang Shih[1]

[1] Department of Physics, The University of Texas at Austin, Austin, TX 78712, USA

[2] Department of Physics and EHSRC, University of Ulsan, Ulsan 680-749, South Korea

[3] Department of Physics, Northeastern University, Boston, MA 02115, USA

[4] Department of Physics, Massachusetts Institute of Technology, Cambridge, MA 02139, USA

**Abstract:**

**Using Indium $\sqrt{7} \times \sqrt{3}$ on Si(111) as an atomically thin superconductor platform, and by systematically controlling the density of nano-hole defects (nanometer size voids), we reveal the impacts of defects density and defects geometric arrangements on superconductivity at macroscopic and microscopic length scales. When nano-hole defects are uniformly dispersed in the atomic layer, the superfluid density monotonically decreases as a function of defect density (from 0.7% to 5% of the surface area) with minor change in the transition temperature *Tc*, measured both microscopically and macroscopically. With a slight increase in the defect density from 5% to 6%, these point defects are organized into defect chains that enclose individual two-dimensional patches. This new geometric arrangement of defects dramatically impacts the superconductivity, leading to the total disappearance of macroscopic superfluid density and the collapse of the microscopic superconducting gap. This study sheds new light on the understanding of how local defects and their geometric arrangement impact superconductivity in the two-dimensional limit.**

Superconducting ground states are known to be robust against non-magnetic disorder [1], in the weakly disordered 3D bulk case. However, in a highly disordered regime, both the transition temperature $T_c$ and the superfluid density (SFD) can be significantly suppressed by disorder induced vortex pinning and scattering centers [2-6]. A conventional superconductor (SC) in the two-dimensional (2D) limit has a low $T_C$ and low SFD, resulting in fragile superconductivity [7-9]. Previous investigations using highly disordered amorphous and granular films have also shown a rapid suppression of both $T_C$ and the SFD with thickness reduction, eventually resulting in a superconductor-insulator transition [3,10-17]. The emergence of single crystal films, however, reveals surprises: at a thickness of only few monolayers, Pb films still show remarkably high superfluid rigidity with robust superconductivity [18-21], indicating the need for a close examination of how superfluid rigidity disappears in single crystal superconducting films in the 2D limit. Intuitively, in the single atomic layer limit, one anticipates that local defects would have a profound impact on superconductivity [1]. But exactly "how" such defects manifest at different length scales in 2D superconductivity remains an unexplored territory. With the rapid discovery of different atomically thin single crystal superconductors [20,22,23], addressing how the defect formation at a microscopic level influences the superconductivity in the 2D limit becomes ever critical and timely.

Using Indium $\sqrt{7} \times \sqrt{3}$ on Si(111) as an atomically thin superconductor platform (Fig. 1a) [24-28], we control the formation of one specific type of defects, nanometer-size hole defects, in terms of density and their geometric arrangements, and investigate the superconductivity from microscopic to macroscopic length scales. Microscopically, we probe the local superconducting gap using scanning tunneling microscopy/spectroscopy (STM/STS) (Fig. 1b) and macroscopically, we probe the SFD using a double coil mutual inductance measurement (Fig. 1c) [2,29-31]. Most significantly, we found that these nano-hole defects have a profound impact on the superconductivity at different length scales. When these nano-hole defects are uniformly dispersed in the 2D film, we found that the SFD decreases monotonically as increasing of defect density while $T_c$, measured both microscopically and macroscopically, remains relatively robust. However, at higher defect concentrations, when these point defects are organized into defect chains that separate the 2D surface into regions of enclosed 2D patches, both the SFD and quasi-particle gap vanish, down to the lowest temperature of our measurements (~ 2.3 K).

The double-coil measures the temperature dependent complex sheet conductivity $Y(T) = [\sigma_1(T) + i\sigma_2(T)]d$, where $d$ is the sample thickness, in our case determined using STM (see supplementary), and $\sigma_1 + i\sigma_2$ is the usual complex conductivity [2,18,29-33]. The real part $\sigma_1$ reflects the dissipative process caused by vortex motion, and the imaginary part $\sigma_2$ is related to the SFD $n_s$, through $\sigma_2 = \frac{n_s e^2}{m\omega}$ [8]. It is customary to refer to $\frac{1}{\lambda^2} = \frac{\mu_0 n_s e^2}{m}$ as the SFD (as they are proportional), and we adopt this convention. This set-up also allows us to directly measure the superfluid phase rigidity $J_s$, through $J_s = \frac{\hbar^2 d}{4e^2 \mu_0 k_B \lambda^2}$. A detailed description of the double-coil set-up and SFD calculation can be found in supplementary. As both the STM/STS and double coil probes are *in situ* and non-contact, the sample crystallinity is maintained and undesirable effects from electrical contact fabrication are avoided in *in situ* transport measurements [24,25,28,34]. By applying these two techniques on the same sample, a direct comparison between microscopic and macroscopic SC behavior can be made.

Starting from a pristine single crystal Indium $\sqrt{7} \times \sqrt{3}$ layer on Si(111), we introduce defects as an independent control parameter (see supplementary). Fig. 2a to 2d show the topography of sample #1 to sample #4 with increasing defects density. The inset atomic images show that all four samples are in the $\sqrt{7} \times \sqrt{3}$ phase. The percentage of the hole defects refers to the surface area fraction occupied by the voids. In addition, the area fraction of extra islands is also labeled. Fig. 2e and 2f show typical zoomed-in images of hole and island defects, both of which are in nanometer scales and cause imperfection on a continuous film. Fig. 2g shows the temperature dependent superfluid density $1/\lambda(T)^2$, for sample #1 to sample #4 and Fig. 2h shows the corresponding real part $\sigma_1$ evolution. Using the two-fluid model fitting, $\frac{1}{\lambda^2}(T) = \frac{1}{\lambda^2}(0\ K)(1 - \frac{T}{T_c})^4$, on sample #1 (Fig. 2g), the zero-temperature SFD can be estimated: $\frac{1}{\lambda^2}(0\ K) = 3.4\ \mu m^{-2}$. From the temperature dependent SFD one can calculate the phase rigidity, $J_s(T)$. Following Emery and Kivelson [9], we evaluate the ratio between the characteristic phase-ordering temperature ($0.9 \times J_s(0\ K)$ for a 2D system) and the superconductivity transition temperature, as it parameterizes the strength and importance of phase fluctuations in the superconductivity transition. The ratio is roughly two for sample #1, indicating a regime where the phase fluctuations play an important role, even for a nearly perfect crystalline film. Note this ratio is markedly different from an earlier study of few-monolayer Pb films whose superfluid rigidity is more than an order of magnitude higher [18], and for bulk Pb, this ratio is more than two orders of magnitude [9].

The temperature dependent SFD in the extreme 2D limit can be described by the BKT theory adapted to the SC scenario [35-38], which is that the universal BKT line with a slope of $\frac{8\pi\mu_0 k_B}{d\Phi_0^2}$ intersects $1/\lambda(T)^2$ at the BKT transition temperature, *i.e.*, $T_{BKT} = \frac{\pi}{2}J_s$. At this temperature thermally excited vortices start to proliferate and destroy the quasi-long-range order. A standard BKT theory would predict a sudden jump in the SFD from zero to a finite value at $T_{BKT}$ [35,37,38].

However, such a sudden jump in the SFD is absent here; instead, the change is gradual, varying across a finite temperature range, suggesting that the behavior here does not follow the traditional BKT theory [39]. Due to this smooth transition, a finite SFD can still be detected above $T_{BKT}$. We define the critical temperature, the onset temperature of detectable SFD as $T_{C,SFD}$. For example, $T_{C,SFD} = 3.3 \pm 0.05\ K$ for sample #1, which is almost the same value as the bulk indium case, 3.4 K. This defined $T_{C,SFD}$ is consistent with the onset temperature of vortex proliferation, which can be seen in the corresponding $\sigma_1(T)$ behavior and is shown later to be consistent with the STS measured transition temperature.

Sample #2 shows similarly low hole density, albeit with a slightly higher island density, compared with sample #1. The SFD result shows a comparable value, although $T_{C,SFD}$ occurs at a slightly lower temperature, $2.95 \pm 0.05\ K$, suggesting that an increase of scattering due to the increased island defects can suppress the $T_C$ slightly but without impacting the SFD. From the two-fluid model fitting, the fitted zero-temperature SFD is 4.2 $\mu m^{-2}$ [40], further testifying that this system is in the strong phase fluctuation limit. Interestingly, the dissipation component, $\sigma_1(T)$ in sample #1 shows a broader width than that in sample #2 despite having a slightly higher $T_C$; this might be related to a slightly larger width of grooves at the step edges in sample #1 which increases the phase fluctuations [41]. A dramatic change occurs in sample #3 when the hole density reaches 5%. Even though 95% of the surface retains its pristine single crystallinity, as shown by the atomic image, the SFD drops by almost one order of magnitude, signaling an enhancement of phase fluctuations. However, the onset SFD temperature $T_{C,SFD}$, is reduced only by 2% and 12%, compared with sample #2 and #1 respectively. This shows that the local hole defects disturb the phase coherence and thus strongly suppress the global phase rigidity but has little effect on $T_{C,SFD}$.

More interestingly, upon a further increase of hole density to 6% (Fig. 2d), we found that the geometric arrangement of defects changes from a uniform distribution to defect chains forming closed loops, which break the continuous film into isolated patches. Although the crystallinity of the atomic structure is still preserved in the flat areas, we can no longer detect SFD down to the lowest instrumentation temperature. This systematic study indicates that both defect density and connectivity profoundly impact the phase rigidity in atomic layer superconductors—a point to be elaborated further below.

We next discuss the local superconducting gap. STS was used to probe the temperature dependent superconducting gap, $\Delta(T)$, using both a normal tip and a superconducting tip; the latter provides higher energy resolution with better accuracy of gap value determination (See Fig. S3). Here, we present detailed results for sample #2 (Fig. 2b and 3a) and sample #3 (Fig. 2c and 3b), where the transition from high to low superfluid phase rigidity occurs. Fig. 3c and 3d show the spectra acquired on sample #2 and #3 using a niobium (Nb) tip and a lead (Pb)-coated tungsten (W) tip respectively, which exhibit SC-SC tunneling features [8], with four peaks at $\pm|\Delta_1 + \Delta_2|$ and $\pm|\Delta_1 - \Delta_2|$, where $\Delta_1$ and $\Delta_2$ refer to the superconducting gaps for tip and sample. A more accurate determination of $\Delta_2$ is based on fitting an SC-SC tunneling formula, and Fig. 3e inset shows one example (see supplementary Fig. S3b for detailed analysis). The BCS fitting of the temperature dependent gap value $\Delta(T)$ (Fig. 3e and 3f) allows us to obtain the transition temperature for sample #2, $T_{C,BCS\_sample\ \#2} = 3.1 \pm 0.1\ K$, and for sample #3, $T_{C,BCS\_sample\ \#3} = 2.9 \pm 0.2\ K$ [42].

Fig. 3g summarize the experimental determination of microscopic $T_{C,BCS}$ (defined by the detectable energy gap) and macroscopic $T_{C,SFD}$ (defined by the detectable SFD) as well as the SFD at 2.3 K for samples of different hole densities. Within the experimental error, we find that the values of $T_{C,BCS}$ are consistent with the values of $T_{C,SFD}$. In addition, in sample #4 where an SFD is not detectable down to 2.3 K, the SC gap is not observed down to 2.3 K either (see supplementary Fig. S4). These results indicate that a macroscopic detectable SFD goes hand-in-hand with a microscopic detectable SC gap. This observation is consistent with that in a 3D conventional superconductor [43], but directly contrasts with highly disordered 2D SC films, where the SC order parameter is spatially non-uniform [14,44]. We think the difference is related to the single-crystal nature of this system, where most of the film is well crystallized and connected, which leads to a uniform pairing potential and a coherent SC transition across the sample. In addition, the significant impact of hole defect density on the SFD is shown in Fig. 3h.

We next discuss whether defects result in an inhomogeneity in the local tunneling gap. Fig. 4b and 4d present the STS mappings across several atomic steps on sample #2 and several defects on sample #3 respectively. Since $\Delta_1$ is the tip SC gap, the spatial uniformity of the sample SC gap $\Delta_2$, is reflected in the uniformity of $\pm|\Delta_1 + \Delta_2|$ peak energies. As sample #2 contains primarily pristine regions, it might not be surprising that the gap uniformity is maintained even across the step edges [41,45]. Most surprisingly, this gap uniformity is maintained on sample #3, which contains 5% hole defects. Outside the gap energy range, the tunneling spectra exhibit a higher noise level at the defect locations. This is further exemplified by the STS spectra (Fig. 4f) acquired at a lower temperature and at three different representative positions (marked on Fig. 4e): the pristine area, the step edge defect, and the hole defect. All three spectra show the typical superconductor to superconductor tunneling features with the same $\pm|\Delta_1 + \Delta_2|$ peak energies, indicating uniform gap values among these three points. Nevertheless, features near $\pm|\Delta_1 - \Delta_2|$ in the defect regions appear to be more smeared out in the defect, suggesting a weakening of order parameter coherence in the defect region without changing the gap size. As for the increased noise level outside the gap energy range, we attribute it to the enhanced local potential fluctuations due to the charging and de-charging process during tunneling [46]. Note that the uniformity of the SC gap is also maintained at 2.9 K, where we observed a vanishing of SC behavior both locally and globally. The reason why the same gap value is measured in the defect region may be closely related to a much longer coherence length, ~ 600 nm for a crystalline film [25,47], which is about two orders of magnitude larger than the defect size. This prevents local defects from disrupting the SC order parameter, although the defects can contribute to the reduction of SFD and the enhancement of phase fluctuations. On the other hand, the SC proximity effect from the surrounding continuous film may also play a role in retaining the SC gap value at the hole defects especially with the circular geometry of the hole defects, which is known to enhance the proximity effect due to an enhanced Andreev reflection [48].

This joint microscopic/macroscopic investigation provides us with a new insight into the role of nano-hole defects on atomically thin 2D superconductors. We show that provided single crystallinity can be maintained over an extended region with very few defects, the Tc can remain relatively high (close to the bulk value) both at microscopic and macroscopic length scales. Nano-hole defects, when uniformly dispersed, can reduce the superfluid density accordingly but with minor change in the Tc value based on the observable SFD at the macroscopic scale and superconducting gap at the microscopic scale. Most intriguingly, when defects form chain structures that break the surfaces into 2D patches with a lateral dimension of 100-200 nm, the superconducting gap and SFD vanish together. We believe this is due to the difficulty in forming a superconducting coherent state in the local region even though pristine single crystallinity is maintained (although the SC state may exist at a much lower temperature). Our work illustrates the profound impacts of nano-hole defects on atomically thin 2D superconductors: both the density and the geometric arrangement of defects disrupt the formation of superconducting states. The overall picture presented here should be relevant to other types of condensates—such as exciton, magnon, and polariton condensates—in the extreme 2D limit.

## Acknowledgements

We are grateful to Allan H. MacDonald, Ming Xie and Takashi Uchihashi for helpful discussions. This work was primarily supported by the National Science Foundation through the Center for Dynamics and Control of Materials: an NSF MRSEC under Cooperative Agreement No. DMR-1720595. Other support was from NSF Grant Nos. DMR-1808751, DMR-1949701, DMR-2114825, and the Welch Foundation F-1672.

# Figures

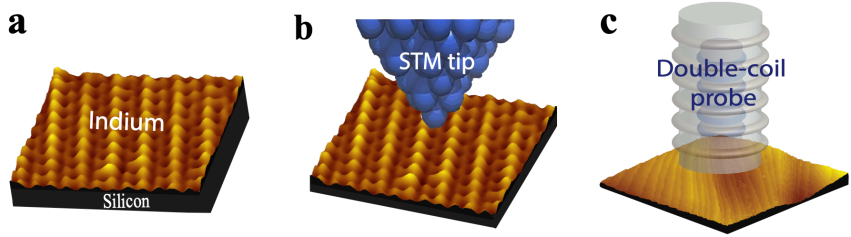

**Fig. 1| Schematic illustration of methodology. a,** Indium adatoms on Si(111) and reconstruct into $\sqrt{7} \times \sqrt{3}$ phase. **b,** Microscopic probe of scanning tunneling microscope. Both the probe apex and the tunneling region are in nm scale. **c,** Macroscopic probe of double-coil mutual inductance system. Both the probe coil size and the sample size are in mm scale.

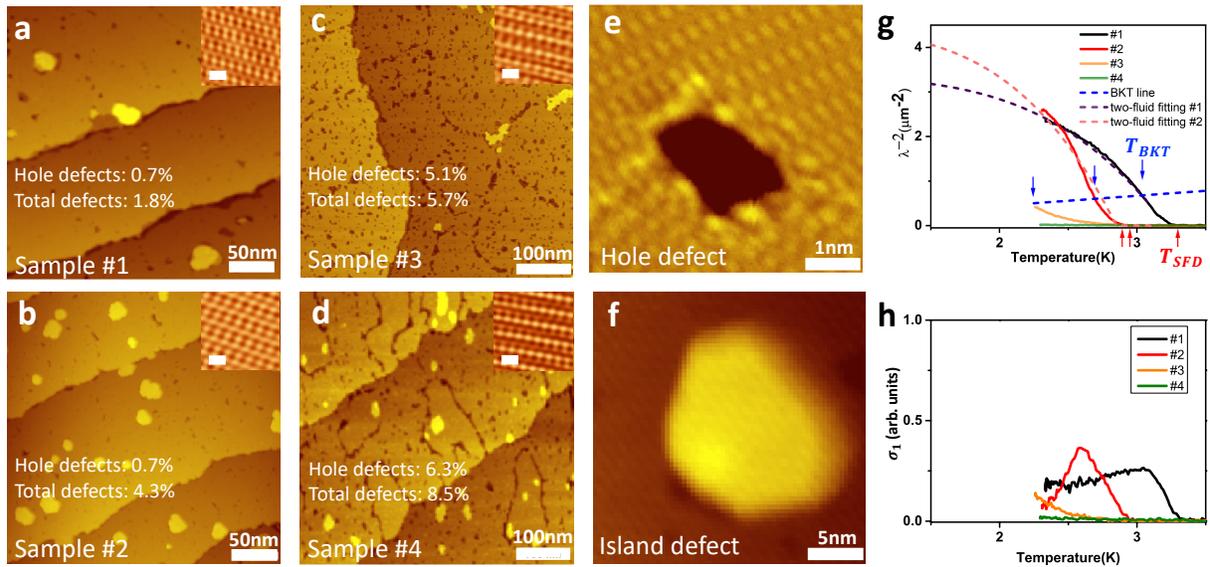

**Fig. 2| Temperature-dependent superfluid density**. **a-d**, Topography of indium $\sqrt{7} \times \sqrt{3}$ monolayer samples with varying defect densities. The top right **insets** show their corresponding atomic images where the scale bar is 1 nm. **e-f**, Topographic image of hole and island defects. **g,** Temperature-dependent superfluid density for sample #1 to #4. The two-fluid model fitting is used

for sample #1 and #2 to extrapolate the SFD at 0 K. The blue dashed line is the universal BKT line. **h,** The temperature-dependent $\sigma_1$ for sample #1 to #4.

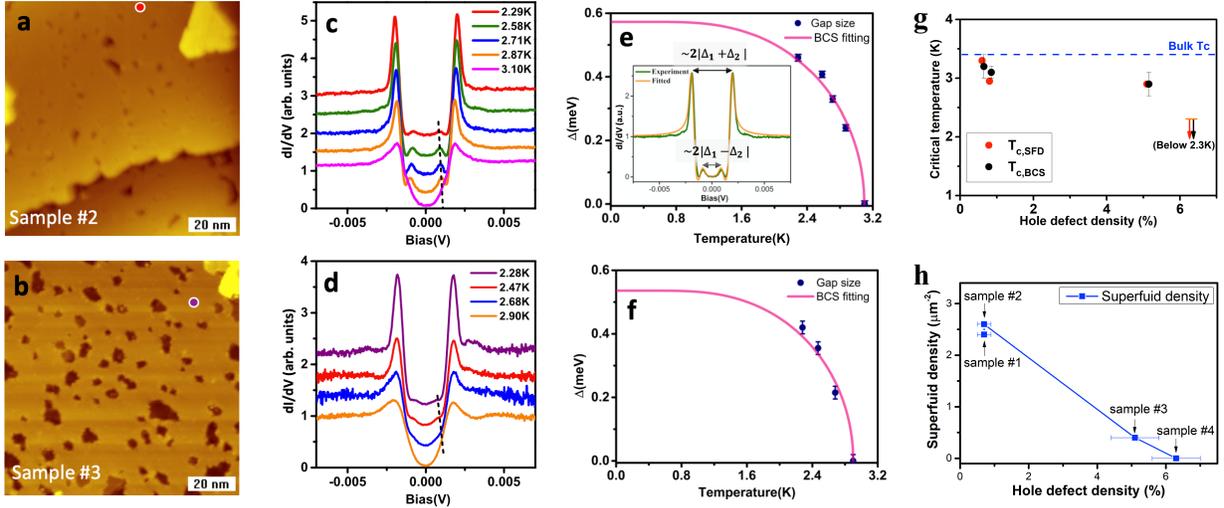

**Fig. 3| Temperature dependent quasi-particle excitation spectrum. a, b,** STM image taken on sample #2 and #3. **c, d,** Temperature dependent tunneling spectra on the pristine area of sample #2 and #3 using superconducting Nb and Pb tips, respectively. Spectra acquisition positions for 2.3 K are marked on Fig. a and b with the corresponding color. Spectra at other temperatures are also taken at a similar area, more than 10 nm away from the hole defects. Curves are offset for clarity. The black dashed line is a guide to show the temperature dependent $|\Delta_1 - \Delta_2|$ tunneling peak position. **e, f,** BCS gap fitting for sample #2 and #3 respectively. The **inset** in **e** shows a typical tunneling spectrum using superconducting Nb tip at 2.58 K. **g,** A summary of critical temperatures $T_{C,SFD}$ and $T_{C,BCS}$ as a function of hole defect density. Data points for sample #1 and #2 are laterally offset to avoid overlapping, both are at 0.7% hole defect concentrations. **h,** A summary of SFD at 2.3 K for samples of different hole defect density. Horizontal error bars represent statistical standard deviations of hole defect density.

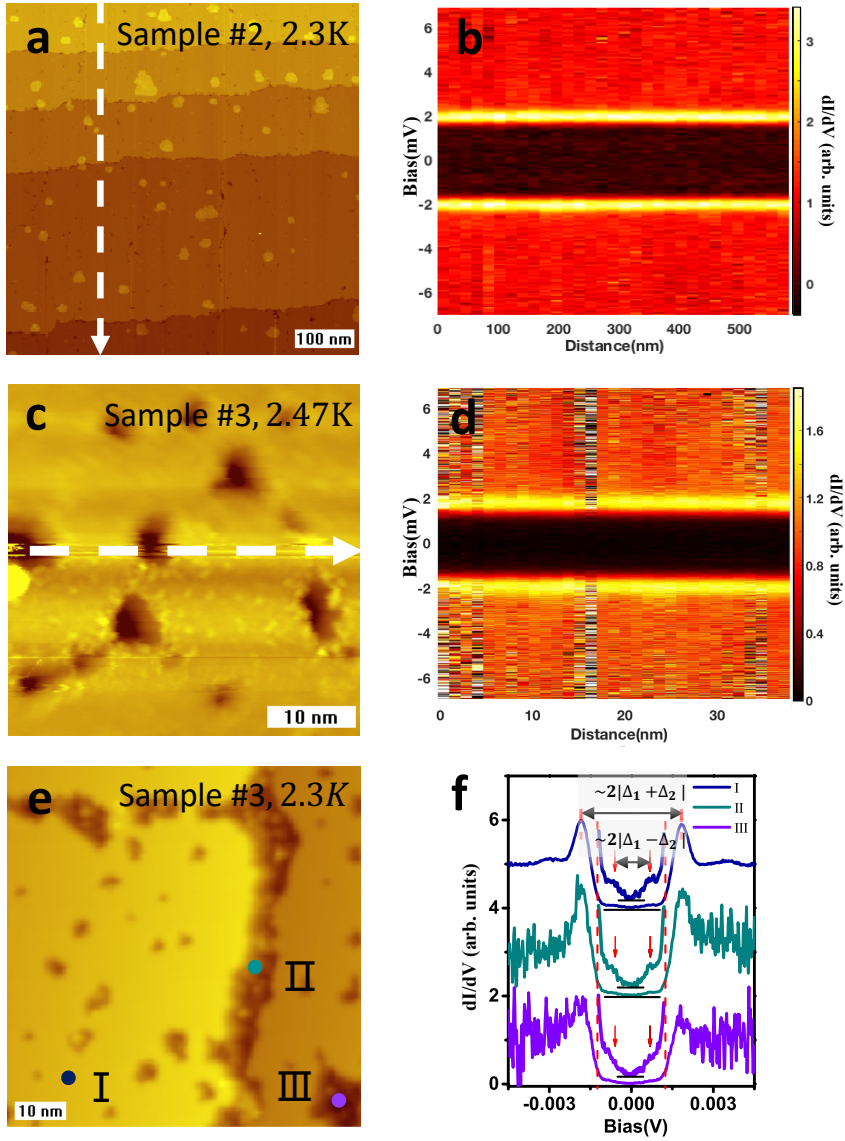

**Fig. 4| Uniformity of superconducting gap distribution. a, c,** STM image taken on sample #2 and #3. **b, d,** The spatial dependence of superconducting gap spectra along the white dashed arrow in **a** and **c** respectively. The position independent $\pm|\Delta_1 + \Delta_2|$ (bright yellow) peak energies show that the gap is spatially uniform. **e,** Topography on sample #3 showing step edge defects and hole defects. **f,** Spectra taken on position I to III as labeled in **e**. Inside the window marked by the red dashed line, duplicate curves which are amplified by a factor of 7 are also plotted to better show

the tunneling peak feature at $\pm|\Delta_1 - \Delta_2|$. Curves are offset for clarity and the horizontal black bars mark the zero for each curve.

# Supplementary information for

# Influence of nano-hole defects and their geometric arrangements on the superfluid density in atomically thin single crystals of indium superconductor

Mengke Liu[1], Hyoungdo Nam[1], Jungdae Kim[1,2], Gregory A. Fiete[3,4], Chih-Kang Shih[1]

[1] Department of Physics, The University of Texas at Austin, Austin, TX 78712, USA

[2] Department of Physics and EHSRC, University of Ulsan, Ulsan 680-749, South Korea

[3] Department of Physics, Northeastern University, Boston, MA 02115, USA

[4] Department of Physics, Massachusetts Institute of Technology, Cambridge, MA 02139, USA

## Part 1. Materials and methods

Our experiments were conducted in home-built STM and non-contact *in situ* double-coil mutual inductance systems, with base pressure to be ~ $10^{-11}$ torr and lowest temperature to be ~ 2.3 K. The indium $\sqrt{7} \times \sqrt{3}$ films were prepared using a home-built molecular beam epitaxy system with the growth pressure to be ~ $10^{-10}$ torr. The nominal miscut angle of the Si substrates is $0 \pm 0.1$ degree. The growth procedure consists of a deposition of 2.3 ML indium onto Si(111) $7 \times 7$ surface (Fig. S1a) at room temperature and *in situ* annealing of the sample for about 5 min.

The density of nano-hole defects is controlled by the annealing temperature. An increase of the annealing temperature away from (higher than) the optimal annealing temperature will induce hole defects, likely due to the evaporation of the indium atoms or some form of de-wetting at a higher temperature. In general, the higher the annealing temperature away from the optimal annealing temperature, the more hole defects there will be, and eventually the film can be evaporated completely leaving only the Si substrate. The annealing temperature for sample #1 to #4 are 225 °C, 230 °C, 235 °C, 240 °C respectively. In addition, we use the thermal radiation heating method (tungsten filament heating), and our annealing stage temperature sensor reading error bar is estimated to be $\pm 5\,°C$. Once the annealing temperature is higher than the optimal annealing temperature (~230 °C), the evaporation rate or de-wetting process of the indium atoms will suddenly increase and increase hole defects to about 5%. It is easier to control the film to be at a relatively well-crystallized and patch-like regime. Fig. S2b to d show additional three samples' STM images: sample #S1 and #S2 both show nice continuity and crystallinity, and sample #S3 shows disconnected patch-like features.

Samples were transferred into STM and double-coil mutual inductance systems through a transfer vessel with $10^{-10}$ torr base pressure in order to maintain the perfect crystallinity of the film. The sample thickness was determined as the depth of the hole, which is $4.2 \pm 0.5\,\text{Å}$. The superconducting niobium (Nb) tip was prepared by mechanical cutting of the Nb wire, followed by *in situ* electron beam cleaning. The Pb-coated W tip was prepared by an electrochemically etched W tip treated with *in situ* electron beam cleaning, and then poked into the Pb films until Pb cluster was transferred onto the tip apex.

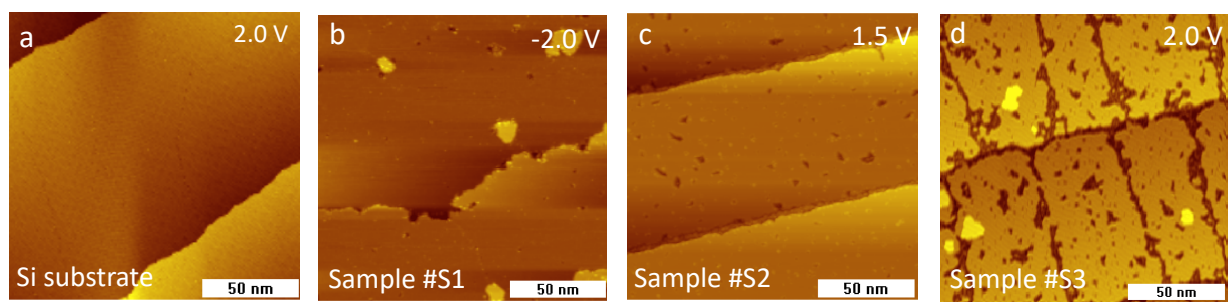

**Fig. S1| Indium film growth protocols. a,** Si(111) substrate. **b-d,** Topographic STM image of sample #S1 to #S3.

**Part 2. Photographic images of instruments**

Fig. S2a is a photographic image of the double-coil mutual inductance system probe [1,2] on top of the sample. A reflection image of the coil probe can be seen on the sample surface and is used to determine the probe distance from the sample surface. In our study, we typically position the probe ∼ 65 μm above the sample, avoiding any mechanical contact. The probe diameter is 1.5 mm, which is smaller than the typical sample size (3.5 mm × 8 mm), to ensure that the response is of macroscopic scale but only resulting from the sample surface. Using this technique, we measure the sample superfluid density as a function of temperature. Fig. S2b is a photographic image for the scanning tunneling microscope (STM) tip on top of a sample where both the STM tip and its reflection image on the sample surface can be seen. The same sample can be transferred through an ultra-high-vacuum transfer vessel with base pressure ∼ $10^{-10}$ torr, between these two probes stations to maintain the sample's perfect crystallinity.

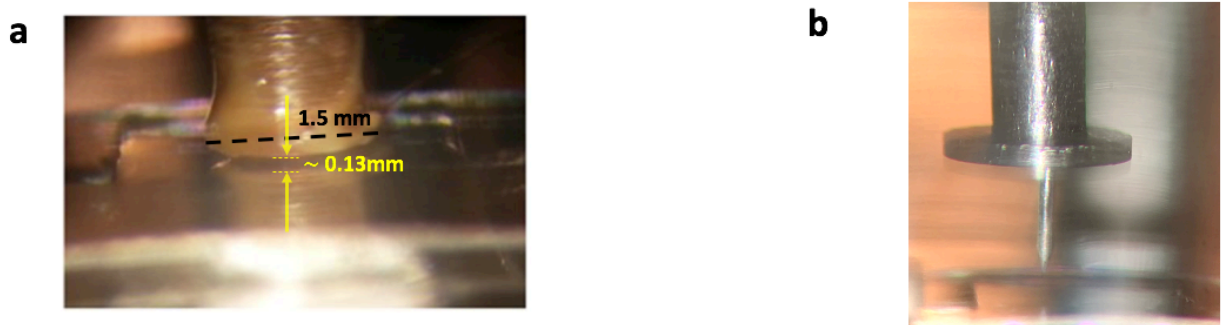

**Figure S2| Photographic images of instruments. a**, Photographic image of the double-coil mutual inductance system probe. The diameter of the probe is about 1.5 mm, and the probe to sample distance is about 65 μm. **b**, Photographic image of the scanning tunneling microscope tip.

**Part 3. Mutual inductance measurement results and superfluid density conversion**

The double-coil mutual inductance probe consists of two concentric coils: driving coil and receiving coil. The AC magnetic field generated by the AC current in the driving coil induces a sheet screening current in the superconducting sample [2-5]. This screening current further leads to an electromotive force in the receiving coil. The relation between the sample sheet conductivity $Y$ and the electromotive force $\delta V$ can be written as [1,6,7]:

$$\delta V = i\omega I_D \int_0^\infty dx \frac{M(x)}{1 - \frac{2}{i\omega\mu_0\eta Y}x}, \quad M(x) = \pi\mu_0\eta\xi(x), \quad Y = [\sigma_1 + i\sigma_2]d$$

$\omega$ represents the driving current frequency, which is 50 kHz and $I_D$ represents the driving current, which is 15 $\mu A$. $\mu_0$ is the vacuum permeability, and $\eta$ is distance between the sample and the coils. $\xi(x)$ is the unitless geometric factor that is related to the detail dimension of our double coil probe and d is the film thickness. The imaginary part of the conductivity is related to the superfluid density through $\sigma_2 = \frac{n_s e^2}{m\omega}$, and $n_s$ is the sample superfluid density [8]. It is a convention to refer to $\frac{1}{\lambda^2} = \frac{\mu_0 n_s e^2}{m}$ as the SFD (as they are proportional), and we adopt this convention. By changing the temperature T, we can gain a temperature dependent SFD, $\frac{1}{\lambda^2}$ (T), of the sample. The error bar of the converted SFD is about 10 percent, which is mainly coming from the sample to probe distance error.

**Part 4. Tunneling spectra using normal tip and superconducting tip**

A comparison between tunneling spectra taken with normal tungsten (W) tip and superconducting niobium (Nb) tip is presented in Fig. S3. The lowest temperature of our STM instrument is ~ 2.3 K, which is close to the superconductivity transition temperature of the indium sample, ~ 3.0 K. Compared with tunneling spectra using normal tip, superconducting tip improves the energy resolution and provides better accuracy of gap size determination. Fig. S3a shows temperature dependent tunneling spectra taken by W tip on indium sample #1. Fig. S3b exhibits an example of tunneling spectra using Nb superconducting tip on indium sample #2 at 2.58K. The spacing between the outer (inner) tunneling peaks roughly corresponds to twice of the sum (difference) of the tip and the sample gap values (as labeled on Fig. S3b); however, an accurate determination of the gap value requires the fitting procedure. The fitting of the experimental data is done through the superconducting to superconducting tunneling formula [8]:

$$I_{ss} = \frac{G_{nn}}{e} \int_{-\infty}^{\infty} \frac{N_{1s}(E)N_{2s}(E+eV)}{N_{1n}(0)N_{2n}(0)}[f(E) - f(E+eV)]dE,$$

$$G_{ss} = \frac{dI_{ss}}{dV}.$$

$I_{ss}$ is the superconductor to superconductor tunneling current; $G_{nn}$ ($G_{ss}$) is the tip to sample tunneling conductance, both tip and sample are at the normal (superconducting) state; $\frac{N_{1s}(E)}{N_{1n}(0)}$ $\left(\frac{N_{2s}(E)}{N_{2n}(0)}\right)$ is the density of states of the first (second) superconductor with respect to its normal state; $f(E)$ is the Fermi-Dirac distribution function. From the fitting, both the sample gap value and the tip gap value are obtained:

$$\Delta_{sample} = 0.408\ meV, \Delta_{tip} = 1.44\ meV.$$

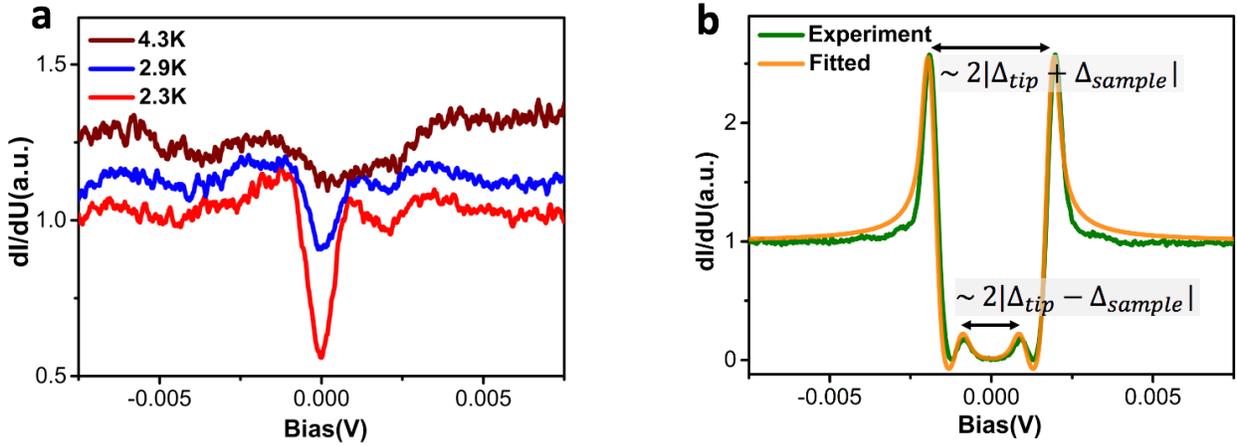

**Figure S3| Tunneling spectra using W tip and Nb tip. a**, Temperature dependent tunneling spectra using W tip taken on indium sample #1. Coherence peaks are broadened because the instrumentation temperature is close to the superconductivity transition temperature. Curves are shifted vertically for clarity. **b**, Tunneling spectra at 2.58K using Nb tip taken on sample #2. The green curve represents the experimental data, and the yellow curve represents the fitted results.

**Part 5. Tunneling spectroscopy on sample #4**

Fig. S4 shows the tunneling spectroscopy on sample #4, using a Pb superconducting tip at 2.3 K. The absence of superconductor-to-superconductor tunneling feature indicates that the superconducting transition temperature for sample #4 is below 2.3 K.

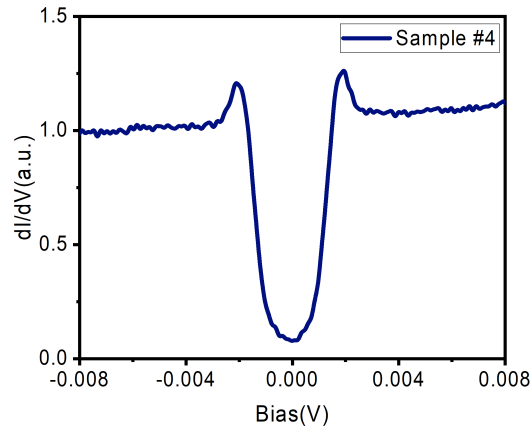

**Fig. S4** | Tunneling spectroscopy on sample #4 using a Pb superconducting tip at 2.3 K.

**Part 6. Additional SFD data analysis**

Fig. S5 shows a plot of "SFD (@T/Tc=0.78) as a function of hole defects density" to take into account the different Tc values between the samples. The trend is similar to that of "SFD (@ T=2.3K) as a function of hole defects density", which further signified the significant impact of hole defect density on superfluid density.

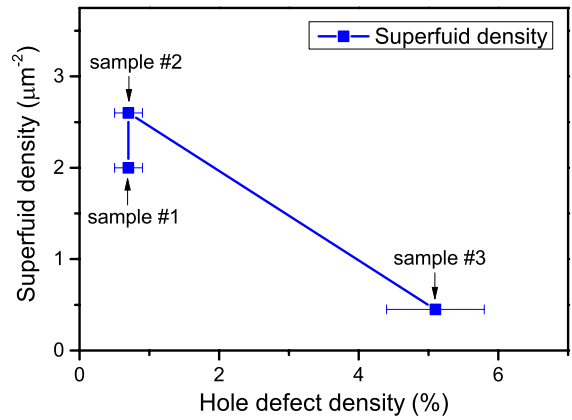

**Fig. S5|** SFD at T/Tc=0.78 for sample #1 to #3.

Fig. S6 shows the double-axis plot of SFD as a function of hole and island defects density, showing that the SFD is strongly correlated with the hole defects density but not with the island defects density.

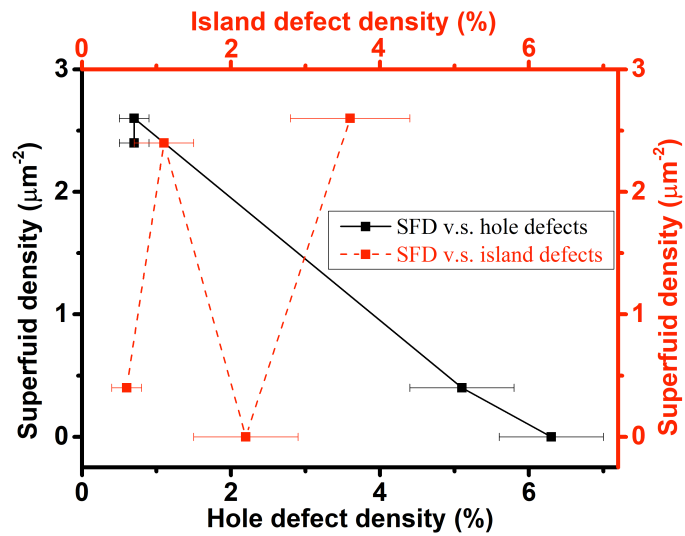

**Fig. S6|** Double-axis plot of superfluid density as a function of hole and island defects density.